# Young's Modulus and Corresponding Orientation in β-Ga$_2$O$_3$ Thin Films Resolved by Nanomechanical Resonators

Xu-Qian Zheng[1], Hongping Zhao[2,3], Zhitai Jia[4], Xutang Tao[4], and Philip X.-L. Feng[1*]

[1]*Department of Electrical and Computer Engineering, Herbert Wertheim College of Engineering, University of Florida, Gainesville, FL 32611, USA*

[2]*Department of Electrical and Computer Engineering, Ohio State University, Columbus, OH 43210, USA*

[3]*Department of Materials Science and Engineering, Ohio State University, Columbus, OH 43210, USA*

[4]*State Key Laboratory of Crystal Materials, Shandong University, Jinan, Shandong 250100, China*

## *Abstract*

We report on the non-destructive measurement of Young's modulus of thin-film single crystal beta gallium oxide (β-Ga$_2$O$_3$) out of its nanoscale mechanical structures by measuring their fundamental mode resonance frequencies. From the measurements, we extract Young's modulus in (100) plane, $E_{Y,(100)}$ = 261.4±20.6 GPa, for β-Ga$_2$O$_3$ nanoflakes synthesized by low-pressure chemical vapor deposition (LPCVD), and Young's modulus in [010] direction, $E_{Y,[010]}$ = 245.8±9.2 GPa, for β-Ga$_2$O$_3$ nanobelts mechanically cleaved from bulk β-Ga$_2$O$_3$ crystal grown by edge-defined film-fed growth (EFG) method. The Young's moduli extracted directly on nanomechanical resonant device platforms are comparable to theoretical values from first-principle calculations and experimentally extracted values from bulk crystal. This study yields important quantitative nanomechanical properties of β-Ga$_2$O$_3$ crystals, and helps pave the way for further engineering of β-Ga$_2$O$_3$ micro/nanoelectromechanical systems (M/NEMS) and transducers.

***Keywords***: β-gallium oxide (β-Ga$_2$O$_3$), Young's modulus, suspended nanostructure, nanomechanics, resonance, micro/nanoelectromechanical systems (M/NEMS)

[*]Corresponding Author. Email: philip.feng@ufl.edu

Beta phase gallium oxide (β-Ga$_2$O$_3$) is the most stable polymorph of gallium oxide at ambient conditions. In addition, the monoclinic crystal structure of β-Ga$_2$O$_3$ is the only crystal structure of Ga$_2$O$_3$ that can be grown from the melt, contributing to its strong potential in cost-effective mass production of high quality, high volume β-Ga$_2$O$_3$ single crystal for wafer scale device fabrication. Intriguingly, β-Ga$_2$O$_3$, as a semiconductor with an ultra-wide bandgap (UWBG, $E_g \approx 4.8$eV),[1,2] also possesses outstanding properties in electrical, optical, and mechanical domains. Thanks to its UWBG, β-Ga$_2$O$_3$ can hold very high electrical field before breaking down (up to $E_{br}$ = 8 MV/cm theoretically).[3,4] This exceptionally high breakdown field engenders strong promises for power electronics and high-voltage radio frequency (RF) applications.[3,4,5] The bandgap at $E_g$ = 4.8 eV also introduces a photon absorption edge at the cut-off wavelength of solar-blind ultraviolet (SBUV) light, making β-Ga$_2$O$_3$ attractive for light detection in SBUV regime.[6,7] Further, β-Ga$_2$O$_3$ has excellent Young's modulus ($E_Y$ = 230–280 GPa according to initial studies),[8,9,10,11,12] which is better than or comparable to those of the conventional materials for micro/nanoelectromechanical systems (M/NEMS), such as silicon ($E_{Y,Si}$ = 130–190 GPa),[13] silicon carbide ($E_{Y,SiC}$ = 390–480 GPa),[14] and gallium nitride ($E_{Y,GaN}$ = 210–405 GPa).[15] Thus β-Ga$_2$O$_3$ is also attractive for M/NEMS applications. On one hand, β-Ga$_2$O$_3$ M/NEMS devices, namely, resonators, oscillators, mixers, filters, *etc.*, can supplement the emerging β-Ga$_2$O$_3$ power and RF electronics.[16,17] On the other hand, β-Ga$_2$O$_3$ M/NEMS can serve as resonant transducers for real-time SBUV detection which can provide better speed and responsivity than their optoelectronic counterparts in certain scenarios.[18,19] To date, the development of β-Ga$_2$O$_3$ M/NEMS is in its infancy and its future improvement relies on gaining new quantitative knowledge and developing comprehensive understanding of mechanical properties of β-Ga$_2$O$_3$ crystals. The Young's modulus of β-Ga$_2$O$_3$ crystal, especially for thin-film β-Ga$_2$O$_3$, has been considered only in a few studies experimentally or theoretically.[8,9,10,11,12] Careful examination and determination of Young's modulus of thin-film β-Ga$_2$O$_3$ in micro/nanoscale structures along with its dependency on the crystal orientation is highly desired for accelerating the development of β-Ga$_2$O$_3$ M/NEMS. Previous studies have suggested that Young's modulus and its corresponding orientation can be resolved from measuring the resonances of properly designed or configured micro/nanomechanical structures.[20,21]

In this Letter, we demonstrate non-destructive Young's modulus extraction for single-crystal β-Ga$_2$O$_3$ thin films by measuring resonance frequencies from two different types of β-Ga$_2$O$_3$ nanomechanical structures produced by different growth methods. The first type of devices (Series A) takes the form of nanodisks clamped at their circular perimeters. The devices are fabricated using β-Ga$_2$O$_3$ nanoflakes grown by low-pressure chemical vapor deposition (LPCVD). The second type of devices (Series B) uses a doubly-clamped beam structure. The beam is constructed by mechanically cleaved β-Ga$_2$O$_3$ nanobelts from bulk crystal synthesized by edge-defined film-fed growth (EFG) method. The devices are carefully selected to ensure the dominance of the flexural rigidity on the resonance frequency so that the initial tension γ [N/m] (or built-in stress σ [N/m$^2$ or Pa]) cannot affect the extraction of Young's modulus (see detailed device design and selection considerations in Supplementary Material). We use ultrasensitive optical interferometry techniques to measure the resonance frequencies of the micro-/nano-mechanical structures.[11] Young's moduli from eight devices are extracted and compared to the values from the literature.

Figure 1 illustrates the crystal structure of β-Ga$_2$O$_3$ which belongs to the monoclinic crystal system, with C2/m space group. Axes *a*, *b*, and *c* define the unit cell of β-Ga$_2$O$_3$ crystal. Axes *a* and *b* are perpendicular to each other, while axis *c* is (*β* =) 103.7° inclined from axis *a*. The unit cell of β-Ga$_2$O$_3$ is anisotropic, where *a* = 12.2 Å, *b* = 5.8 Å, and *c* = 3.0 Å. To simplify the representation of directions in the crystal, we define an orthogonal coordinate system with respect to the unit cell of β-Ga$_2$O$_3$. The $x$ and $y$ axes coincide with $a$ ([100] direction) and $b$ ([010] direction) axes of β-Ga$_2$O$_3$ crystal, respectively. Therefore, we have $z$ axis perpendicular to $x$ and $y$ axes, which is 13.7° off from $c$ axis as shown in Fig. 1a. Such directions can also correspond to the crystal orientations in nanostructures of Series A and Series B devices as shown in Fig. 1b and Fig. 1c, respectively.

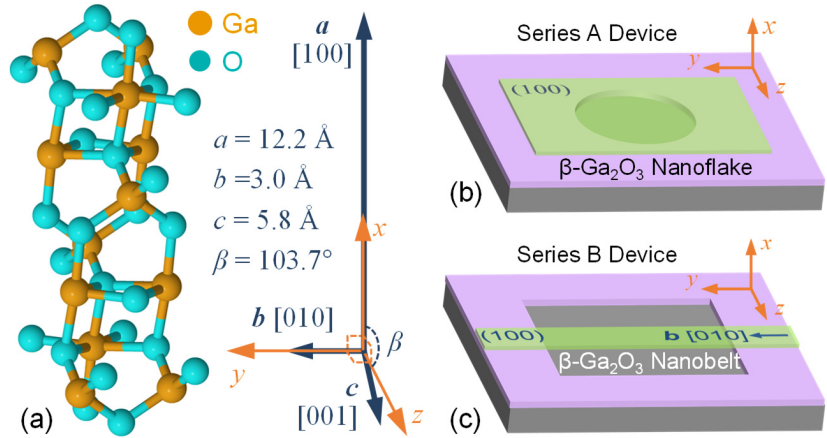

**FIG. 1.** (a) Schematic of monoclinic structure of β-Ga$_2$O$_3$ crystal with coordinative illustration of the relations between $x$, $y$, $z$ directions and crystal orientations of β-Ga$_2$O$_3$. Illustration of (b) Series A devices and (c) Series B devices.

The first series (Series A) of β-Ga$_2$O$_3$ resonators are fabricated using β-Ga$_2$O$_3$ nanoflakes synthesized by LPCVD method on a 3C-SiC-on-Si substrate.[22] The devices in Series A are selected from the devices used in Ref. 11. Using high purity Ga pellets and O$_2$ (Ar as carrier gas) as source materials, β-Ga$_2$O$_3$ is grown on a 3C-SiC epi-layer on Si substrate at a growth temperature of 950 °C for 1.5 hours. The as-grown β-Ga$_2$O$_3$ is in the form of nanoflakes. We use thermal release tape to pick up the β-Ga$_2$O$_3$ nanoflakes and apply the nanoflakes to a 290nm-SiO$_2$-on-Si substrate with predefined microtrenches. Then, the structure is heated up to 90 °C. By gently removing the thermal release tape, the β-Ga$_2$O$_3$ nanoflakes are deposited on the substrate. The deposited β-Ga$_2$O$_3$ nanoflakes along with the pre-defined circular microtrenches form suspended nanostructures in the form of nanodisks clamped at the peripheral of the circular microtrenches.

We conduct electron backscatter diffraction (EBSD) measurement on multiple deposited β-Ga$_2$O$_3$ nanoflakes (Fig. 2a). The EBSD pole figures (Fig. 2b) suggest that the major surfaces of the β-Ga$_2$O$_3$ nanoflakes are in parallel with the (100) plane of the crystal (perpendicular to the $x$ direction in Fig. 1), while $y$ (or [010]) and $z$ directions are in parallel with the substrate plane. The EBSD results also confirm that the as-grown β-Ga$_2$O$_3$ nanoflakes are single crystal, or at least polycrystalline with only very few grains (Fig. 2c). Further, if there is a grain boundary, it is clearly shown on the flake (as labeled in Fig. 2a). Therefore, by making sure no grain boundary runs through the suspended region, we can guarantee that the suspended part of Series A resonators is single crystalline (as it is always within one grain).

Figure 3a shows a typical β-Ga$_2$O$_3$ disk (Device A1) suspended over a circular microtrench with a diameter of $d$ = 5.24 μm. EBSD measurement suggests that the β-Ga$_2$O$_3$ flake has its $x$ axis pointing into the substrate (Fig. 3b), while the $y$ and $z$ directions are along the edges of the β-Ga$_2$O$_3$ flake. Using atomic force microscopy (AFM), we determine the thickness of the device as $h = 61$ nm (Fig. 3c). By measuring the thermomechanical noise resonance frequency of Device A1, we resolve a fundamental mode resonance at a frequency of $f_0 = 39.6$ MHz with a quality ($Q$) factor of $Q = 626$. When the resonance frequency of such disk device is dominated by the flexural rigidity of the suspended structure, *i.e.*, the device is in the 'disk' (or 'plate') regime, the averaged Young's modulus in device plane can be revealed from the measured fundamental mode resonance frequency $f_0$ by[11,23]

$$E_Y = \frac{3\pi^2 \rho (1-v^2) d^4}{(k_{A0} r)^4 h^2} f_0^2, \qquad (1)$$

where $\rho = 5950$ kg/m$^3$ is the mass density of β-Ga$_2$O$_3$, $v = 0.2$ is Poisson's ratio, and $k_{A0} r = 3.196$ is the eigenvalue for the fundamental mode of a disk resonator clamped at the circular perimeter. Using Eq. (1), we extract a Young's modulus, $E_Y \approx 292$ GPa, of Device A1.

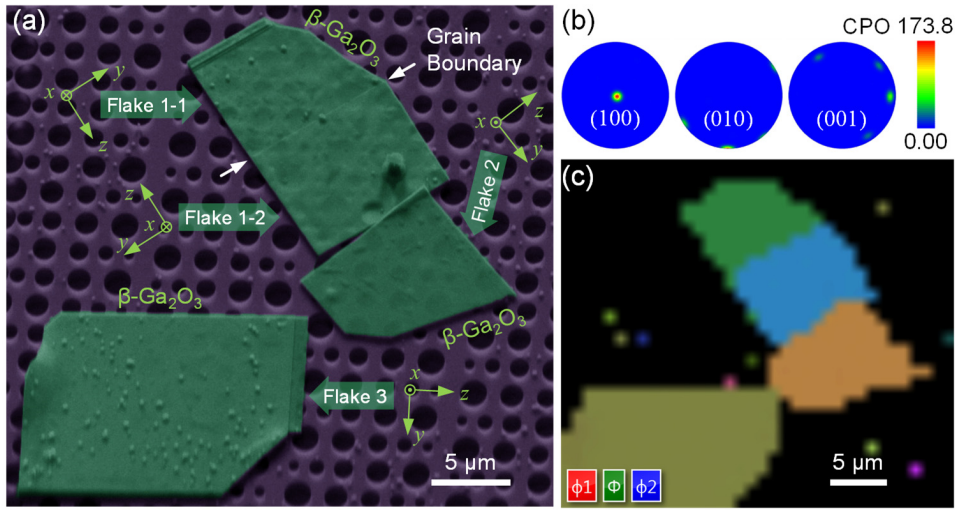

**FIG. 2.** EBSD results of β-Ga$_2$O$_3$ flakes. (a) An SEM image of β-Ga$_2$O$_3$ nanoflakes on substrate with circular microtrenches, with corresponding orientations of each flake labeled in *x-y-z* coordinates. (b) EBSD pole figure for (100), (010), and (001) planes of the β-Ga$_2$O$_3$ flakes in (a). CPO: crystallographic preferred orientation. (c) Corresponding Euler color plot from EBSD measurement for flakes in panel (a).

To further validate the extracted Young's modulus, we measure more β-Ga$_2$O$_3$ disk resonators (see detailed device design and selection considerations in Supplementary Material). Figures 3e & 3f show thermomechanical noise spectra of two β-Ga$_2$O$_3$ disk resonators, Device A2 and Device A3, respectively. We measure altogether five β-Ga$_2$O$_3$ disk resonators with thickness in 39 to 73 nm range, diameters among 3.20–5.24 μm, and fundamental mode resonance frequencies ranging from 39.6 to 74.9 MHz. Thus, we extract an averaged Young's modulus of $E_{Y,(100)} = 261.4 \pm 20.6$ GPa in (100) plane of β-Ga$_2$O$_3$ flakes grown by LPCVD method. Table I summarizes the sizes, resonance frequencies, and extracted Young's moduli of all Series A devices. Since fundamental mode resonance frequency of β-Ga$_2$O$_3$ disk scales linearly with the geometry ratio $h/d^2$ of the disk (*inset* of Fig. 3g, modified from Eq. (1)), we can plot the frequency scaling with respect to $h/d^2$ using $E_Y = 261.4$ GPa (Fig. 3g). All the measured data points are close to the curve predicted by Eq. (1), indicating great precision in extraction of Young's modulus using the resonance frequencies of the β-Ga$_2$O$_3$ nanodisk resonators.

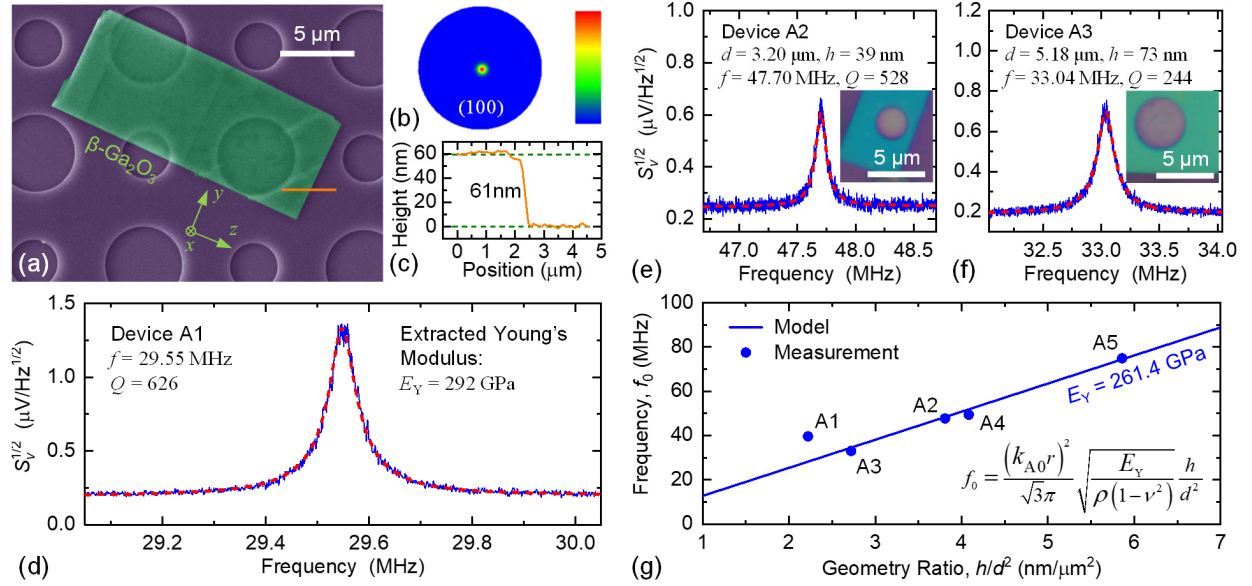

**FIG. 3.** Young's modulus extraction from β-Ga$_2$O$_3$ disk resonators. (a) A colored SEM image of a β-Ga$_2$O$_3$ disk resonator (Device A1). Coordinate labels indicate $x$, $y$, and $z$ directions acquired from EBSD measurement. (b) EBSD pole figure for (100) plane of Device A1. (c) AFM trace corresponding to the orange line in (a). (d) Fundamental mode resonance spectrum of Device A1. (e) & (f) Fundamental mode resonance spectra of Devices A2 and A3, respectively. *Insets*: device images. (g) Resonance frequency scaling with respect to geometry ratio $h/d^2$ using the extracted Young's modulus ($E_Y$ = 261.4±20.6 GPa) for LPCVD β-Ga$_2$O$_3$ crystal.

TABLE I. Extracted Young's Moduli from β-Ga$_2$O$_3$ Disk Resonators

| Device Number | Diameter $d$ (μm) | Thickness $h$ (nm) | Resonance Frequency $f_0$ (MHz) | Young's Modulus $E_Y$ (GPa) |
|---|---|---|---|---|
| A1 | 5.24 | 61 | 39.6 | 292 |
| A2 | 3.20 | 39 | 47.7 | 259 |
| A3 | 5.18 | 73 | 33.0 | 244 |
| A4 | 3.20 | 60 | 74.9 | 270 |
| A5 | 3.32 | 45 | 49.4 | 242 |
| Average | | | | 261.4±20.6 |

We use the β-Ga$_2$O$_3$ nanobelts that are mechanically cleaved from bulk β-Ga$_2$O$_3$ crystal grown by EFG method for fabrication of second series (Series B) of β-Ga$_2$O$_3$ devices.[24] During the fabrication, we first introduce bulk β-Ga$_2$O$_3$ crystal onto a piece of tape. After exfoliating the β-Ga$_2$O$_3$ for multiple iterations using the tape, similar to exfoliation of two-dimensional (2D) materials,[25] we apply the exfoliated β-Ga$_2$O$_3$ flakes onto a polydimethylsiloxane (PDMS) stamp for transfer. Through observation under optical microscope, the exfoliated β-Ga$_2$O$_3$ flakes are in the form of nanobelts. Based on the bonding strength of different planes of monoclinic β-Ga$_2$O$_3$ crystal, the largest surface of the β-Ga$_2$O$_3$ nanobelt is in parallel with the (100) plane of β-Ga$_2$O$_3$ crystal and the [010] direction ($y$ axis) goes along with the longest sides of the nanobelt.[24,26] After identifying the β-Ga$_2$O$_3$ nanobelt with desired size and thickness, we transfer the nanobelt on PDMS to a 290 nm-SiO$_2$-on-Si substrate with pre-defined rectangular microtrenches using the dry transfer technique.[25] The β-Ga$_2$O$_3$ nanobelt suspended over the microtrench forms a doubly-clamped beam structure. Figure 4a shows a typical β-Ga$_2$O$_3$ doubly-clamped beam resonator (Device B1). The device is 460 nm thick, suspended over a 20.6 μm long microtrench. The *x*, *y*, and *z* directions of β-Ga$_2$O$_3$ crystal are labeled in Fig. 4a based on the orientation of the β-Ga$_2$O$_3$ nanobelt. We fabricate and measure three β-Ga$_2$O$_3$ doubly-clamped beams. Device images of Devices B2 and B3 are shown in Figs. 4c & 4e, respectively. Figures 4b, 4d & 4f show the photothermally driven resonance spectra of the fundamental mode resonances of Devices B1, B2, and B3, respectively. Similar to the β-Ga$_2$O$_3$ disk resonators, Young's modulus of the doubly-clamped beam can be extracted from measured first mode resonance frequency when the resonance motion of the device is dominated by the flexural rigidity of the structure. Therefore, we have[27]

$$E_Y = \frac{48\pi^2 \rho L^4}{(k_{B0}L)^4 h^2} f_0^2, \qquad (2)$$

where $L$ is the length of the doubly-clamped beam and $k_{B0}L = 4.73$ is the eigenvalue for the fundamental mode of a doubly-clamped beam. In this case, we can extract Young's modulus along the [010] direction (*y* axis) of β-Ga$_2$O$_3$ crystal. Table II summarizes the sizes, resonance frequencies, and extracted Young's moduli of all Series B devices. We have an averaged Young's modulus of $E_Y = 245.8 \pm 9.2$ GPa along [010] direction (*y* axis). Similar to β-Ga$_2$O$_3$ disk resonators, the fundamental mode resonance frequencies of β-Ga$_2$O$_3$ doubly-clamped beams scale linearly with the geometry ratio $h/L^2$ (*inset* of Fig. 4g, modified from Eq. (2)). Thus we can also confirm the precision of Young's modulus extraction for doubly-clamped beams with the results in Fig. 4g.

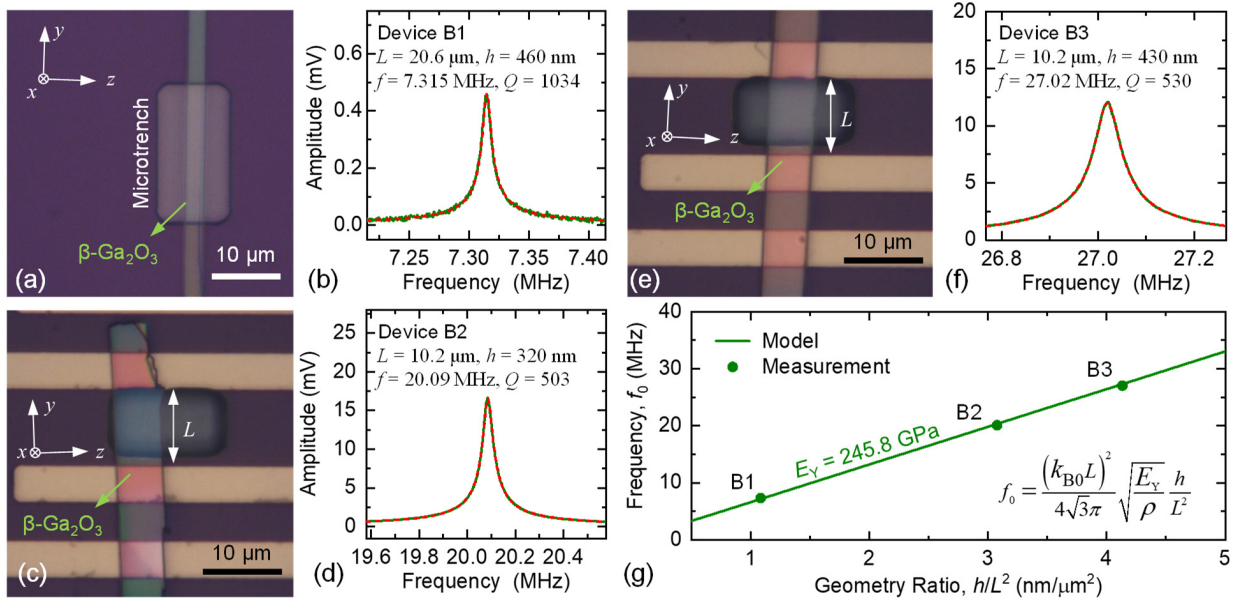

**FIG. 4.** Young's modulus extraction from β-Ga$_2$O$_3$ doubly-clamped beam resonators. (a), (c) & (e) Optical images of β-Ga$_2$O$_3$ doubly-clamped beam resonators, Devices B1, B2, and B3, respectively. Coordinate labels indicate *x*, *y*, and *z* directions based on the orientation of the β-Ga$_2$O$_3$ nanobelt. The suspension length $L$ of Devices B2 and B3 are confined by the electrodes. (b), (d) & (f) Fundamental mode resonance spectra of Devices B1, B2, and B3, respectively. (g) Frequency scaling with respect to geometry ratio $h/L^2$ using the extracted Young's modulus ($E_Y = 245.8 \pm 9.2$ GPa) for EFG β-Ga$_2$O$_3$ crystal.

TABLE II. Extracted Young's Moduli from β-Ga$_2$O$_3$ Doubly-Clamped Beam Resonators

| Device Number | Length $L$ (μm) | Thickness $h$ (nm) | Resonance Frequency $f_0$ (MHz) | Young's Modulus $E_Y$ (GPa) |
|---|---|---|---|---|
| B1 | 20.6 | 460 | 7.315 | 256.4 |
| B2 | 10.2 | 320 | 20.09 | 240.3 |
| B3 | 10.2 | 430 | 27.02 | 240.7 |
| Average | | | | 245.8±9.2 |

Utilizing the measured fundamental mode resonance frequencies, we have extracted the Young's modulus in (100) plane (from Series A devices), $E_{Y,(100)}$ = 261.4±20.6 GPa, for LPCVD grown β-Ga$_2$O$_3$ nanoflakes, and in [010] direction (y axis, from Series B devices), $E_{Y,[010]}$ = 245.8±9.2 GPa, for exfoliated nanobelts from bulk crystal synthesized by EFG method, respectively. The values are comparable to measured Young's modulus values of β-Ga$_2$O$_3$ in other structures, e.g., $E_Y$ = 232 GPa for (100) plane of bulk β-Ga$_2$O$_3$ (Ref.10) and $E_Y$ = 280 GPa for nanowire.[8] While lower Young's moduli ($E_Y$ = 120–215 GPa) have been reported in nanoindentation tests of polycrystalline β-Ga$_2$O$_3$ thin films grown on SiC buffer layers on Si substrates with crystallographic planes of (001), (011) and (111),[12] crystal orientations of β-Ga$_2$O$_3$ are not determined and correlation with $E_Y$ values cannot be established. The monoclinic crystal structure of single-crystal β-Ga$_2$O$_3$ gives rise to possibly anisotropic Young's modulus. Using the theoretically calculated elasticity (stiffness) tensor matrices of β-Ga$_2$O$_3$ in Refs. 9 and 12, we can calculate the Young's modulus in (100) plane (y-z plane, Fig. 5) by using the equations in Ref. 28. The Young's modulus in this plane is much larger in y (or [010]) and z directions. Correspondingly, Young's moduli in y (or [010]) and z directions are $E_{Y,y}$ = 329.9 GPa and $E_{Y,z}$ = 297.6 GPa, and $E_{Y,y}$ = 262.2 GPa and $E_{Y,z}$ = 222.9 GPa, from theoretical stiffness matrices in Ref. 9 and Ref. 12, respectively. Similarly, Young's moduli in y (or [010]) and z directions are $E_{Y,y}$ = 287.7 GPa and $E_{Y,z}$ = 243.7 GPa, and $E_{Y,y}$ = 275.6 GPa and $E_{Y,z}$ = 241.3 GPa, from experimentally measured stiffness matrices using resonant ultrasound spectroscopy (RUS) from bulk β-Ga$_2$O$_3$ crystals in Ref. 9 and Ref. 29, respectively. The measured Young's modulus in (100) plane (solid blue circular curve) and Young's modulus in y (or [010]) direction (green stars) in this work are comparable with projected theoretical and bulk crystal values in Fig. 5. To further compare our measured Young's modulus in (100) plane ($E_Y$ = 261.4 GPa) to data converted from the computed and measured stiffness matrices from bulk crystals, we first compute the fundamental-mode flexural resonance frequency of a β-Ga$_2$O$_3$ disk ($d$ = 3 μm, $h$ = 40 nm, similar to Device A2 in Table I) by COMSOL Multiphysics simulation using the stiffness matrix $C$ from Ref. 12, it gives $f_0$ = 51.715 MHz. If we assume an isotropic Young's modulus with the measured value in (100) plane ($E_Y$ = 261.4 GPa), the simulation yields $f_{0,meas}$ = 56.468 MHz. Next, we rescale the stiffness matrix by the factor $(f_{0,meas}/f_0)^2$, and generate a new stiffness matrix $C_{rescaled} = (f_{0,meas}/f_0)^2 C$ = $(56.468/51.715)^2 C$, which produces $f_0$ = 56.460 MHz from simulation. We can then compute the projected anisotropic Young's moduli in (100) plane by using $C_{rescaled}$ based on measurement results from our Series A devices, plotted as the blue dashed line in Fig. 5, which lies between the theoretical results from Ref. 9 (magenta dash-dot-dot line) and Ref. 12 (orange dot line) and is slightly larger than experimental results (dark cyan dash-dot line) converted from stiffness matrix measured by RUS from bulk β-Ga$_2$O$_3$ in Ref. 9. We note that (100) plane is one of the most important main orientations of wafers synthesized by the Czochralski (CZ) method,[30] and [010] is the direction at which β-Ga$_2$O$_3$ has the highest thermal conductivity, $\kappa_{[010]}$ = 27 W/(m K), compared to $\kappa_{[100]}$ = 11 W/(m K) and $\kappa_{[001]}$ = 15 W/(m K) (Ref. 31). By measuring the resonance frequencies of different nanomechanical structures in this study, the Young's moduli of β-Ga$_2$O$_3$ corresponding to these important crystal orientations have been extracted non-destructively from the nanomechanical device platforms, which are directly applicable to future design and development of β-Ga$_2$O$_3$ M/NEMS and other mechanically coupled or tuned devices.

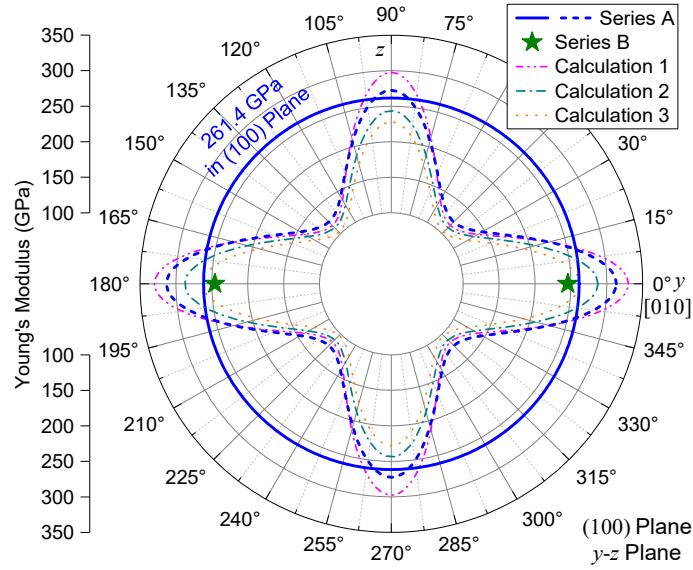

**FIG. 5.** Projection and comparison of Young's modulus of β-Ga$_2$O$_3$ in the (100) plane (or *y-z* plane). Symbols and solid line are measured data from this work. The blue dashed line is our calculation and projection of Young's modulus in (100) plane based on experimental results from our Series A devices, by employing theoretical models from Ref. 28. Magenta dash-dot-dot line (Calculation 1) shows our calculated projection of Young's modulus in (100) plane using the theoretical stiffness matrix from Ref. 9; dark cyan dash-dot line (Calculation 2) displays the projected Young's modulus using stiffness matrix measured from bulk crystal in Ref. 9; orange dot line (Calculation 3) depicts projected Young's modulus using the theoretical stiffness matrix from Ref. 12. We have employed equations from Ref. 28 in all the above $E_Y$ calculations using different stiffness matrix data.

In conclusion, we have demonstrated the extraction of Young's modulus in (100) plane, $E_{Y,(100)}$ = 261.4±20.6 GPa, for LPCVD grown β-Ga$_2$O$_3$ nanoflakes (thickness 39–73nm) and Young's modulus in [010] direction, $E_{Y,[010]}$ = 245.8±9.2 GPa, of 320–460nm-thick β-Ga$_2$O$_3$ nanobelts exfoliated from EFG synthesized β-Ga$_2$O$_3$ bulk crystal, using the measured fundamental flexural mode resonance frequencies of β-Ga$_2$O$_3$ nanomechanical structures. These extracted Young's moduli are comparable to those obtained from bulk crystals experimentally or from theoretical calculations. This work provides essential information regarding mechanical properties in micro/nanoscale β-Ga$_2$O$_3$ crystals for future design and development of β-Ga$_2$O$_3$ M/NEMS devices for power electronics, RF, and optical sensing applications.

**Data Availability Statement:** Data available in article.

**Supplementary Material:** See Supplementary Material for theoretical analysis and the device design and selection considerations of β-Ga$_2$O$_3$ nanomechanical resonators.

**Acknowledgement:** X.-Q. Zheng and P. X.-L. Feng thank the financial support from the Defense Threat Reduction Agency (DTRA) Basic Scientific Research Program (Grant No. HDTRA1-19-1-0035). H. Zhao thanks the financial support from the National Science Foundation (NSF) Division of Materials Research (DMR) (Grant No. 1755479). Z. Jia and X. Tao thank the financial support from the Ministry of Science and Technology of the People's Republic of China and the 111 Project 2.0 (Grant No. BP2018013)

# References

[1] M. R. Lorenz, J. F. Woods, and R. J. Gambino, "Some Electrical Properties of the Semiconductor β-$Ga_2O_3$," *J. Phys. Chem. Solids* **28**, 403-404 (1967).

[2] N. Ueda, H. Hosono, R. Waseda, and H. Kawazoe, "Anisotropy of Electrical and Optical Properties in β-$Ga_2O_3$ Single Crystals," *Appl. Phys. Lett.* **71**, 933 (1997).

[3] M. Higashiwaki, K. Sasaki, A. Kuramata, *et al.*, "Gallium Oxide ($Ga_2O_3$) Metal-Semiconductor Field-Effect Transistors on Single-Crystal β-$Ga_2O_3$ (010) Substrates," *Appl. Phys. Lett.* **100**, 013504 (2012).

[4] M. Higashiwaki, K. Sasaki, H. Murakami, *et al.*, "Recent Progress in $Ga_2O_3$ Power Devices," *Semicond. Sci. Technol.* **31**, 034001 (2016).

[5] A. J. Green, K. D. Chabak, M. Baldini, *et al.*, "β-$Ga_2O_3$ MOSFETs for Radio Frequency Operation," *IEEE Electron Device Lett.* **38**, 790-793 (2017).

[6] W.-Y. Kong, G.-A. Wu, K.-Y. Wang, *et al.*, "Graphene-β-$Ga_2O_3$ Heterojunction for Highly Sensitive Deep UV Photodetector Application," *Adv. Mater.* **28**, 10725- 10731 (2016).

[7] R. Lin, W. Zheng, D. Zhang, *et al.*, "High-Performance Graphene/β-$Ga_2O_3$ Heterojunction Deep-Ultraviolet Photodetector with Hot-Electron Excited Carrier Multiplication," *ACS Appl. Mater. Interfaces* **10**, 22419-22426 (2018).

[8] M.-F. Yu, M. Z. Atashbar, and X. Chen, "Mechanical and Electrical Characterization of β-$Ga_2O_3$ Nanostructures for Sensing Applications," *IEEE Sens. J.* **5**, 20-25 (2005).

[9] W. Miller, K. Böttcher, Z. Galazka, and J. Schreuer, "Numerical modelling of the Czochralski growth of β-$Ga_2O_3$," *Cryst.* **7**, 26 (2017).

[10] V.I. Nikolaev, V. Maslov, S.I. Stepanov, *et al.*, "Growth and Characterization of β-$Ga_2O_3$ Crystals," *J. Cryst. Growth* **457**, 132-136 (2017).

[11] X.-Q. Zheng, J. Lee, S. Rafique, *et al.*, "Ultrawide Band Gap β-$Ga_2O_3$ Nanomechanical Resonators with Spatially Visualized Multimode Motion," *ACS Appl. Mater. Interfaces* **9**, 43090-43097 (2017).

[12] A. S. Grashchenko, S. A. Kukushkin, V. I. Nikolaev, *et al.*, "Study of the Anisotropic Elastoplastic Properties of β-$Ga_2O_3$ Films synthesized on SiC/Si Substrates," *Phys. Solid State* **60**, 852-857 (2018).

[13] M. A. Hopcroft, W. D. Nix, and T. W. Kenny, "What is the Young's Modulus of Silicon?" *J. Microelectromech. Syst.* **19**, 229-238 (2010).

[14] H. Chen, H. Jia, C. A. Zorman, *et al.*, "Determination of Elastic Modulus of Silicon Carbide (SiC) Thin Diaphragms via Mode-Dependent Duffing Nonlinear Resonances," *J. Microelectromech. Syst.* **29**, 783-789 (2020).

[15] M. Rais-Zadeh, V. J. Gokhale, A. Ansari, *et al.*, "Gallium Nitride as an Electromechanical Material," *J. Microelectromech. Syst.* **23**, 1252-1271 (2014).

[16] X.-Q. Zheng, J. Lee, and P. X.-L. Feng, "Beta Gallium Oxide (β-Ga$_2$O$_3$) Vibrating Channel Transistor," in Proc. IEEE 33rd Int. Conf. on Micro Electro Mechanical Systems (MEMS), Vancouver, BC, Canada, 18-22 Jan. 2020, pp. 186-189.

[17] X.-Q. Zheng, T. Kaisar, and P. X.-L. Feng, "Electromechanical Coupling and Motion Transduction in β-Ga$_2$O$_3$ Vibrating Channel Transistors," *Appl. Phys. Lett.* **117**, 243504 (2020).

[18] X.-Q. Zheng, J. Lee, S. Rafique, *et al.*, "β-Ga$_2$O$_3$ NEMS Oscillator for Real-Time Middle Ultraviolet (MUV) Light Detection," *IEEE Electron Device Lett.* **39**, 1230-1233 (2018).

[19] X.-Q. Zheng, Y. Xie, J. Lee, *et al.*, "Beta Gallium Oxide (β-Ga$_2$O$_3$) Nanoelectromechanical Transducer for Dual-Modality Solar-Blind Ultraviolet Light Detection," *APL Mater.* **7**, 022523 (2019).

[20] E. J. Boyd and D. Uttamchandani, "Measurement of the Anisotropy of Young's Modulus in Single-Crystal Silicon," *J. Microelectromech. Syst.* **21**, 243-249 (2012).

[21] Z. Wang, H. Jia, X.-Q. Zheng, *et al.*, "Resolving and Tuning Mechanical Anisotropy in Black Phosphorus via Nanomechanical Multimode Resonance Spectromicroscopy," *Nano Lett.* **16**, 5394-5400 (2016).

[22] S. Rafique, L. Han, J. Lee, *et al.*, "Synthesis and Characterization of Ga$_2$O$_3$ Nanosheets on 3C-SiC-on-Si by Low Pressure Chemical Vapor Deposition," *J. Vac. Sci. Technol. B* **35**, 011208 (2017).

[23] H. Suzuki, N. Yamaguchi, and H. Izumi, "Theoretical and Experimental Studies on the Resonance Frequencies of a Stretched Circular Plate: Application to Japanese Drum Diaphragms," *Acoust. Sci. Technol.* **30**, 348-354 (2009).

[24] W. Mu, Z. Jia, Y. Yin, *et al.*, "One-Step Exfoliation of Ultra-Smooth β-Ga$_2$O$_3$ Wafers from Bulk Crystal for Photodetectors," *CrystEngComm* **19**, 5122-5127 (2017).

[25] R. Yang, X.-Q. Zheng, Z. Wang, *et al.*, "Multilayer MoS$_2$ Transistors Enabled by a Facile Dry-Transfer Technique and Thermal Annealing," *J. Vac. Sci. Technol. B* **32**, 061203 (2014).

[26] W. S. Hwang, A. Verma, H. Peelaers, *et al.*, "High-Voltage Field Effect Transistors with Wide-Bandgap β-Ga$_2$O$_3$ Nanomembranes," *Appl. Phys. Lett.* **104**, 203111 (2014).

[27] S. Timoshenko, D. H. Young, and W. Weaver, Vibration Problems in Engineering, Forth Edition (John Wiley & Sons, New York, 1974).

[28] T. C. T. Ting, "The stationary Values of Young's Modulus for Monoclinic and Triclinic Materials," *J. Mech.* **21**, 249-253 (2005).

[29] K. Adachi, H. Ogi, N. Takeuchi, *et al.*, "Unusual Elasticity of Monoclinic β-Ga$_2$O$_3$," *J. Appl. Phys.* **124**, 085102 (2018).

[30] Z. Galazka, S. Ganschow, K. Irmscher, *et al.*, "Bulk Single Crystals of β-Ga$_2$O$_3$ and Ga-Based Spinels as Ultra-Wide Bandgap Transparent Semiconducting Oxides," *Prog. Cryst. Growth Charact. Mater.* **67**, 100511 (2021).

[31] Z. Guo, A. Verma, X. Wu, *et al.*, "Anisotropic Thermal Conductivity in Single Crystal β-Gallium Oxide," *Appl. Phys. Lett.* **106**, 111909 (2015).

*– Supplementary Material –*

# Young's Modulus and Corresponding Orientation in β-Ga$_2$O$_3$ Thin Films Resolved by Nanomechanical Resonators

Xu-Qian Zheng[1], Hongping Zhao[2,3], Zhitai Jia[4], Xutang Tao[4], and Philip X.-L. Feng[1*]

[1]*Department of Electrical and Computer Engineering, Herbert Wertheim College of Engineering, University of Florida, Gainesville, FL 32611, USA*

[2]*Department of Electrical and Computer Engineering, Ohio State University, Columbus, OH 43210, USA*

[3]*Department of Materials Science and Engineering, Ohio State University, Columbus, OH 43210, USA*

[4]*State Key Laboratory of Crystal Materials, Shandong University, Jinan, Shandong 250100, China*

## Table of Contents

---

[*]Corresponding Author.  Email:  philip.feng@ufl.edu

## S1. Device Design and Selection Considerations

In certain devices, the initial tension (or built-in stress) can play significant or even dominant role in determining the resonance frequency of the resonator. Therefore, the nanomechanical resonators in this work are carefully selected (larger thickness-to-diameter/length ratio) to exclude the effect of initial tension (or built-in stress) on the devices. Please find the analysis below.

In a given circular drumhead device, both flexural rigidity $D$ [N m] ($D = E_Y h^3/[12(1-v^2)]$, where $E_Y$ is the Young's modulus [Pa or N/m$^2$], $h$ is the thickness of the device, and $v$ is the Poisson's ratio. Here $D$ is dominated by thickness and Young's modulus) and initial tension $\gamma$ [N/m] can be important. When only flexural rigidity $D$ is considered, we have the equation shown in Fig. 3g of the Main Text:

$$f_0 = \frac{(k_{A0}r)^2}{\sqrt{3}\pi}\sqrt{\frac{E_Y}{\rho(1-v^2)}}\frac{h}{d^2}, \tag{S1}$$

where $(k_{A0}r) = 3.196$ is the eigenvalue, $r$ is the radius of circular resonator, $\rho$ is the volume mass. We can add the tension effect in, and we have[1,2]

$$f_0 = \frac{(k_{A0}r)^2}{\sqrt{3}\pi}\sqrt{\frac{E_Y}{\rho(1-v^2)}}\frac{h}{d^2}\sqrt{1+\frac{3(1-v^2)\gamma d^2}{(k_{A0}r)^2 E_Y h^3}}. \tag{S2}$$

The terms under second square root introduce the tension effect. As shown in Fig. S1a, by using typical range of surface tension values ($\gamma = 0.02$–$2$ N/m) in such devices and plotting the frequency scaling by using Eq. (S2), the resonance frequencies of the devices used for Young's modulus extraction (solid symbols) is in the range that the surface tension (built-in stress) has minimal effect. Rather, it follows the disk regime model (dimmed straight lines plotted using Eq. (S1)). Thus, the Young's modulus extraction from these devices using the resonance frequency is immune from the effect of typical initial tension in Series A devices.

Similarly, when only flexural rigidity is considered, we can estimate the resonance frequency of first mode of the doubly-clamped structure using the equation in Fig. 4g of the Main Text:

$$f_0 = \frac{(k_{B0}L)^2}{4\sqrt{3}\pi}\sqrt{\frac{E_Y}{\rho}}\frac{h}{L^2}, \tag{S3}$$

where $L$ is the length of the doubly-clamped beam and $k_{B0}L = 4.73$ is the eigenvalue. By adding the effects from built-in stress, we have the equation:[3]

$$f_0 = \frac{(k_{B0}L)^2}{4\sqrt{3}\pi}\sqrt{\frac{E_Y}{\rho}}\frac{h}{L^2}\sqrt{1+\frac{2.91\sigma L^2}{\pi^2 E_Y h^2}}, \tag{S4}$$

where $\sigma$ [Pa or N/m$^2$] is the built-in stress of the suspended structure. Using a typical range of built-in stress ($\sigma = 2$–$50$ MPa), we can plot the frequency scaling using Eq. (S4) in Fig. S1b. The resonances of measured devices lie in the regime dominated by the flexural rigidity and follow the dimmed straight lines plotted using Eq. (S3). Therefore, the built-in stress should not affect the Young's modulus extraction based on resonance frequencies from Series B devices.

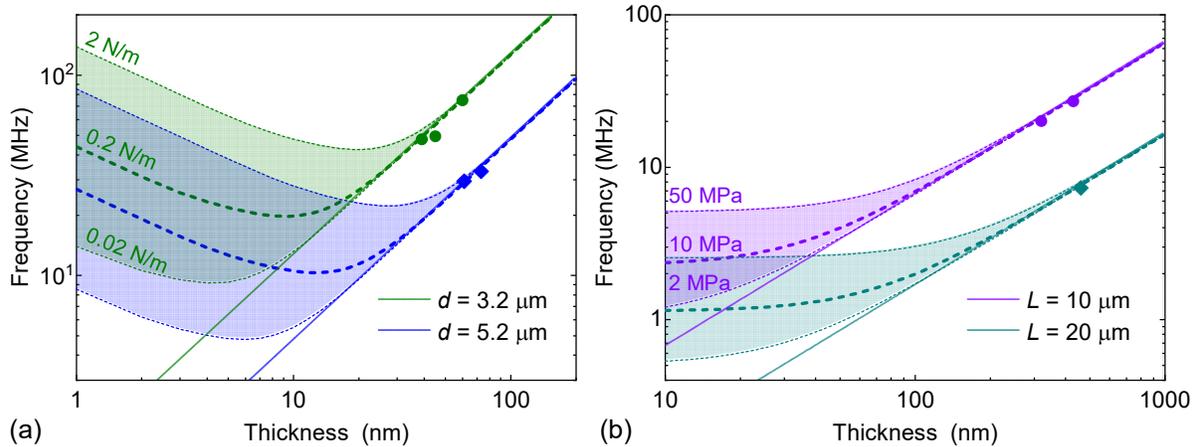

**Figure S1.** Comparison of modeled frequency scaling and measured resonance frequencies of (a) circular (Series A) resonators and (b) doubly-clamped (Series B) resonators. Dashed lines represent modeled frequencies with initial tension or built-in stress considered. Dimmed straight lines represent modeled frequencies with dominance of flexural rigidity. Solid symbols represent the measured fundamental mode resonance frequencies.

## References

[1] H. Suzuki, N. Yamaguchi, and H. Izumi, "Theoretical and Experimental Studies on the Resonance Frequencies of a Stretched Circular Plate: Application to Japanese Drum Diaphragms", *Acoust. Sci. Technol.* **30**, 348-354 (2009).

[2] X.-Q. Zheng, J. Lee, S. Rafique, *et al.*, "Ultrawide Band Gap β-$Ga_2O_3$ Nanomechanical Resonators with Spatially Visualized Multimode Motion", *ACS Appl. Mater. Interfaces* **9**, 43090-43097 (2017).

[3] A. Bokaian, "Natural Frequencies of Beams under Tensile Axial Loads", *J. Sound Vib.* **142**, 481-498 (1990).